# Reversible Gates in Emerging Quantum-dot Cellular Automata Technology: An Innovative Approach to Design and Simulation

Moein Sarvaghad-Moghaddam[1], Ali A. Orouji[1], Zeinab Ramezani[2], Iraj Sadegh Amiri[3,4*],
Alireza Mahdavi Nejad[5]

[1]*Department of Electrical and Computer Engineering, Semnan University, Semnan, Iran.*
[2] *Department of Electrical and Computer Engineering, Northeastern University, Boston, MA 02115, USA.*
[3] *Computational Optics Research Group, Advanced Institute of Materials Science,*
*Ton Duc Thang University, Ho Chi Minh City, Vietnam.*
[4] *Faculty of Applied Sciences, Ton Duc Thang University, Ho Chi Minh City, Vietnam*
[5] *Department of Interdisciplinary Engineering, Wentworth Institute of Technology, Boston, MA 02115, USA*
Emails: moeinsarvaghad@aut.ac.ir, aliaorouji@semnan.ac.ir, Z.ramezani@northeastern.edu,
irajsadeghamiri@tdtu.edu.vn, Mahdavinejada@wit.edu,

**Abstract-** Power dissipation is known as the most notable limiting factor in all nano-electronic design techniques including Quantum-dot Cellular Automata (QCA). The familiar reversible computing approach is used as a reasonably reliable solution, mitigating power dissipation. This study presents, a comprehensive multi-objective method for designing R-Fs in emerging QCA technology. The results are investigated in both logical and layout levels, in detail. The results verify that the approach offered in this study has advantage over the most efficient approaches available in the literature by far. This comparison can be made on various parameters ranging from area, complexity (cell amount), delay (clocking zones), and to even logical levels including levels, Control inputs, the number of majority and NOT gates.

## 1. Introduction

Nanotechnology researches have been focused on overcoming to the physical restrictions of current CMOS scaling at the end of the roadmap [1]. Energy dissipation is the main obstacle in the progress of nano-scale computational systems [2]. Reversible computing is a remedy to this issue. Reversible logic (RL) became important when Landauer found the existence of a lower theoretical limit of energy consumption in computation. Therefore, irreversible functions (IR-Fs) lead to information loss and energy dissipation [3]. Energy dissipates by losing every bit of information quantified as $KTln2$ where $K$ is the Boltzmann constant and $T$ is the absolute temperature of the environment [4]. At room temperature ($T = 27 °C$), the energy dissipation is estimated to $2.9 \times 10^{-21}$ Joules which seems a very small value whereas it is significant in nano-scale circuits where operation under very low power consumption is needed. Bennett proved when the circuit contains reversible gates (RGs), the energy dissipation can be zero [4].

The reduction of feature sizes in designing CMOS is necessary by continuing advances in semiconductor technology. Physical restrictions including quantum effects and non-deterministic behaviors related to small currents work against scaling of feature sizes. An alternatives to CMOS technology is a Quantum-dot Cellular Automata (QCA) [5-9] in which the configuration of an electron pair within a quantum-dot cell identifies its logical state. Lent introduced the pioneering concept of QCA (1993) [10, 11]. QCA cell is the basic unit of a QCA which in turn includes 4 dots locating symmetrically at the square's corners. The QCA technology works based on electrostatic Coulomb interactions between electric charges (e.g. electrons) enclosed within a quantum dot (Q). The clock is undertaken to control the information flow and synchronize it provided power to run the circuit, in QCA Logic [5, 12]. In QCA, the 3-input majority gate (MG) and NOT gate (NG) are the fundamental elements.

Lent et al. introduced the concept of employing QCA in a RL in the design process of an electric circuit [13]; Hereafter many researches offered a plenty of several RGs in QCA [14-17]. In [14], QCA has been investigated for testable implementations of RL. Also, two new RGs (referred to as QCA1 and QCA2) have been proposed for QCA implementation. In [17], a new approach to synthesize a reversible universal QCA logic gate (RUG) structure has been proposed to decrease the garbage outputs (GOs) as well as the logic gates of a design. Also, the RUG gate had been utizied for the realization of QCA logic circuits. In [18] a QCA layout design for the proposed Toffoli gate has been presented. Sarker and et al. [19] have proposed a novel design of a QCA Peres Gate (PG) and its simulation. Bahar et. al. introduced an efficient design for a Fredkin gate [20]. They compared their results with the available ones in literature where the number of cells, covered area, and latency time are used for comparison. Mohammadi et. al. [21] investigated a 1-bit full adder design employing a QCA implementation of Toffoli and Fredkin gates. Right after, a full-adder design with reversible QCA1 gates has been proposed regarding the overhead and power savings. The output function of QCA1 is expressed as:
$Y_1 = x_1 x_2 + x_2 x_3 + x_1 x_3, Y_2 = x_1 x_2 + x_1 x_3' + x_2 x_3', Y_3 = x_1' x_2 + x_1' x_3 + x_2 x_3$.

In this paper, the main contribution is classified in three different parts. In the first part, a multi-objective synthesis method is presented for reversible functions (R-Fs) by extending the synthesis method proposed in [22]. In this part, contol inputs which are the input cells with fixed polarization are used for programming 2-input OR and AND gates. The reduction of control inputs is an achivement used in this paper to improve the reduction of power consumption in R-F. Based on this method, a new synthesis for popular RGs such as Toffoli, Fredkin and so many others are proposed. In the second part, an innovative optimized approach applicable in QCA technology is presented. This novel approach optimally converts an IR-F to a R-F with no need to employ any intermediate scheme. This approach has advantages including direct and optimal conversion of an irreversible function to its counterpart reversible function. The method introduced in this study does not need to call for any optimization method as a subroutine. Therefore, the ongoing vital need of employing any conventional reversible block (RB) (e.g. Toffoli and Fredkin) is eliminated. Because of the minimum number of garbage outputs (GOs), the GOs do not need to generate any extra gates (e.g. MG and NG). This method is applied to the 13 standard combinational functions proposed in [23], and a reversible function is created with the lowest number of MG and GOs. This process is performed to estimate the efficiency of our proposed method. Finally, in the third part, new designs of QCA layouts are presented for gates synthesized in the previous section. Results show that our proposed method outperforms the most powerful methods in terms of area, complexity (cell amount), delay (clocking zones), and in logic level with respect to gate levels, control inputs, number of MG and NG.

## 2. Background material

This part reviews the fundamental concepts in QCA technology with the emphasis on Quantum-dot cellular automata and QCA devices.

### 2.1. Quantum-dot cellular automata

A standard QCA consists of four quantum-dot with two extra electrons all are located symmetrically in the four corners of a square as illustrated in Fig. 1.a. They occupy diagonal antipodal sites through tunneling junctions with Coulombic interaction between the electronic charges. Due to the existence of large potential barriers, it is supposed that tunneling out of the cell is not possible. As shown in Fig. 1.b), a single QCA cell can accommodate two completely polarized "cell polarization" states: binary '1' state ($P = +1$) and binary '0' state ($P = -1$).

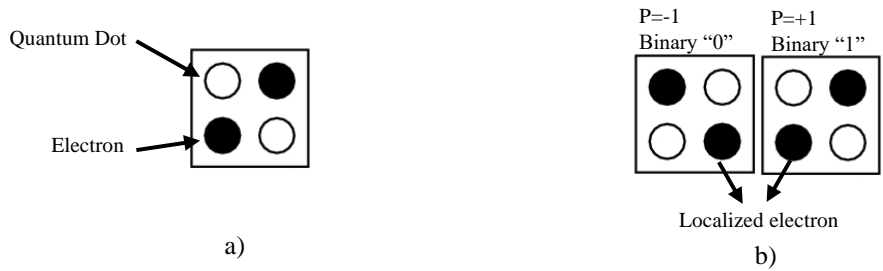

Fig. 1: Illustration of a QCA cell with a) four Qs. b) two different polarizations.

There are four manners for the QCA implementation reported so far include of Semiconductor QCA [24, 25], Magnetic QCA [26-28], Molecular QCA [29, 30], and Metal Dot QCA [31-33]. Orlov et al [34] first have introduced fabrication of the metal dot QCA cell as shown in Fig. 2.a). It consists of four metal connected by $Al - AlO_x - Al$ tunnel junctions. The electron beam lithography (e-beam lithography) and Shadow evaporation techniques both at the absolute temperature of 15mK are used to fabricate the QCA cell. A magnetic field intensity of 1T is required to generate superconductivity of the metal. Electron exchanges between successive dots, resulting in the creation of polarization change. To switch from one dot to the other one, Gate electrodes force tunneling an electron. In the molecular manner [35], three allyl groups are used as connected in a "V" like structure by alkyl bridges. Fig. 2.b) show the structure of molecular QCA cell in states '0', '1' and NULL. In semiconductor manner [7], in such silicon materials as GaAs/AlGaAs QCA cell can be implemented as well as four Qs which can have a high mobility 2D electron gas as shown in Fig.2c which is below the surface. As presented in references [36, 37], and shown in Fig.2.d, a single nanomagnet which has a single magnetization state viz. up (↑) (binary '1' state) or down (↓) (binary '0' state) has been utilized and has turned as magnetic island considering the magnetic manner. By employing the EBL and standard lift-off technique, the magnetic islands can be fabricated in the ranges of 30-nm thin film of permalloy, therefore the nano-sized ferromagnetic materials can form the QCA cells.

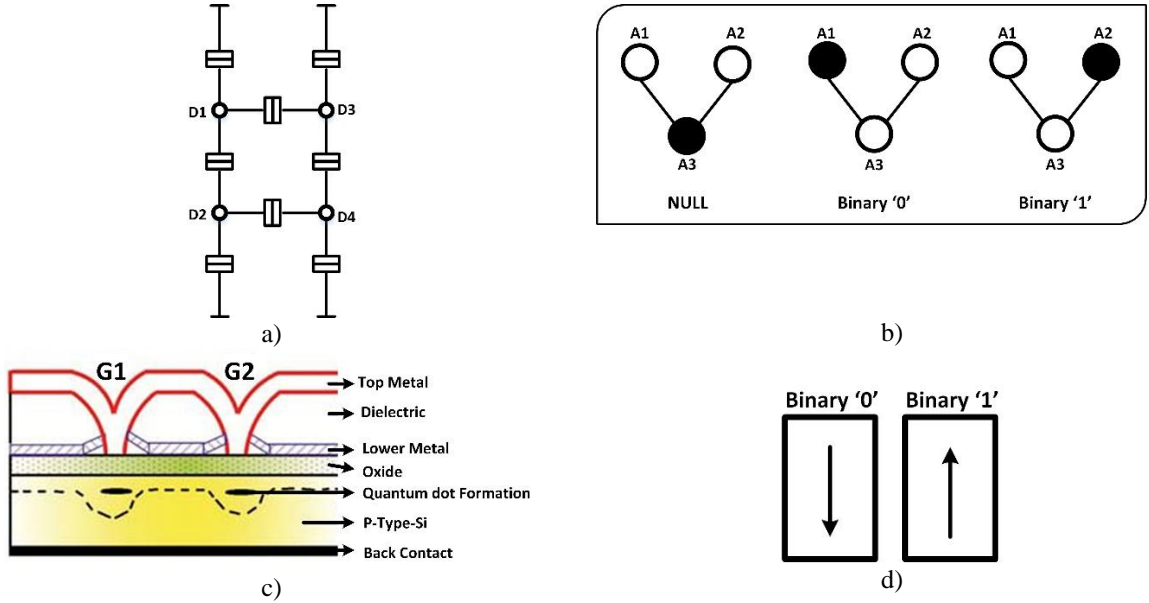

Fig. 2: Illustration of QCA implementation in four different manners. a) Schematic of metal-dot QCA. b) Charge configuration of the molecular QCA, hole as positive charge has been presented by a filled circle transferred between different dots and creates three different states ('NULL', 'Binary 0' and 'Binary 1' respectively). c) Schematic model of semiconductor QCA. d) Binary states '0' and '1' in a Single-domain nanomagnetic QCA cell.

### 2.2. QCA Logic devices

QCA wire, MG, and inverter are the fundamental QCA logic devices. A line of $90°$ QCA cells forms a QCA wire. The QCA wire is inserted to the input cell by another cell using the polarization which is fixed/held [38]. The generated signal which is excited from the farthest left cell spreads along the QCA wire from left to right. Also, a QCA wire can be created by cells rotated by $45°$. This type of wire spreads the input signal and its inversion in odd and even cells, respectively [39, 40]. A QCA MG present a 3-input logic function as written in (1) in which A, B, and C are arbitrary inputs.

$$M(A, B, C) = AB + BC + CA. \tag{1}$$

By imposing one out of 3 inputs of the MG to a fixed logic "0" or a "1" the MG can be utilized to make AND/OR operations as written in the below equations:

$$M(A, B, 0) = AB, \qquad M(A, B, 1) = A + B. \tag{2}$$

Fig. 3 shows a QCA wire in two different angles of $90°$ (part a), $45°$ (part b), inverter gate (part c), and MG (part d).

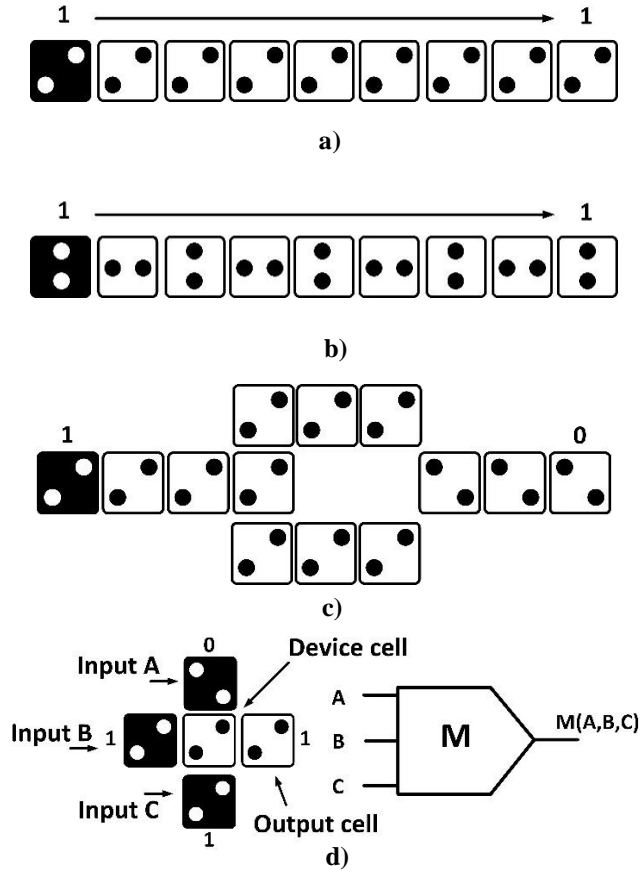

Fig. 3: Illustration of a) 90° and b) 45° QCA wires, c) inverter gate d) QCA MG

Co-planer and multi-layer crossovers are known as two possible crossover options. In first approach, one quantum wire by a 45-degree passes over a conventional (C-) quantum wire without any interposition (Figure 4.a). On the other, multi-layered structures are utilized to pass the quantum wires over each other (Figure 4.b).

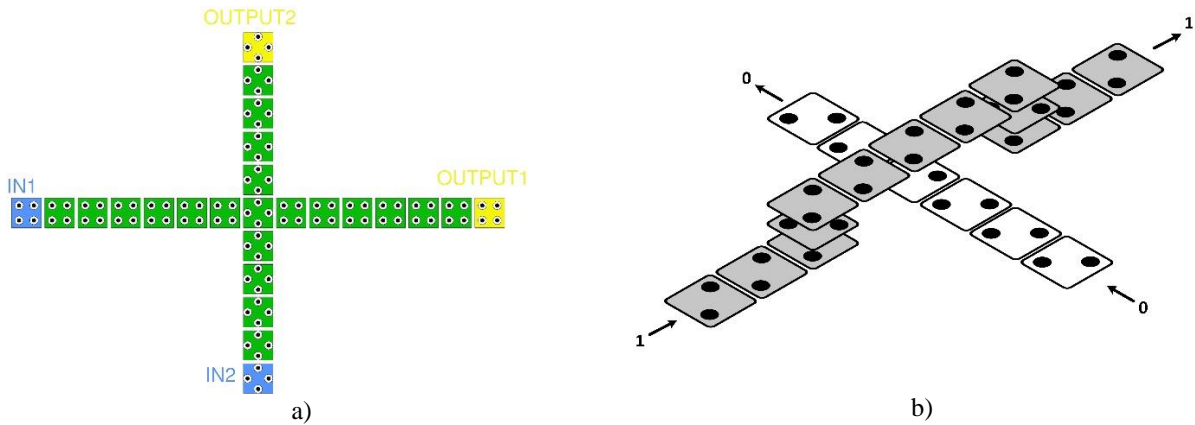

Fig. 4: a) co-planar b) multi-layer crossing wire

### 2.3. QCA clocking

The process of erasure and storage of information inside the cell is controlled by a clock in QCA. For synchronize and information flow control in QCA circuit, Clocking is required. Two following strategies have been introduced to clock a QCA [41]: 1) Landauer clocking: has a logically invariable "erase" operation 2) Bennett clocking: has a logically variable "copy-then-erase" operation [2]. It has already been proved that eliminate without any duplication dissipates energy in the order of magnitude of the signal energy.

Commonly, 4 multi-phase clocking signals of phase lagging of $\frac{\pi}{2}$ are used as presented in Fig. 5.a). This type of clocking system is called Landauer type [42]. During a complete cycle, each zone goes through the four phases. These zones must follow a particular order viz. $C_0 \rightarrow C_1 \rightarrow C_2 \rightarrow C_3$. These four phases corresponds to a complete cycle are named Switch, Hold, Release, and Relax, respectively[5]. Fig. 5(b) shows waveform of Bennett clocking. In this figure some abbreviation used which are switch phase (S), hold phase(H), release phase (Re), and relax phase (Rx).

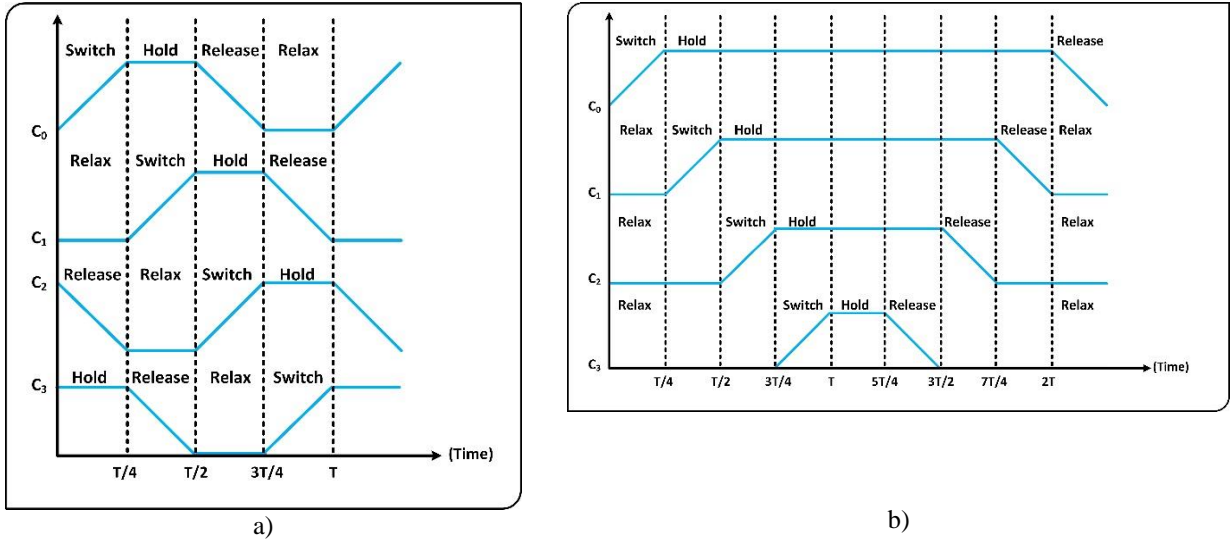

a)            b)

Fig. 5: a) and b) Laundauer and Bennett clocking waveform. Rx: relax phase, Re: release phase, H: hold phase, S: switch phase,

### 2.4. Simulation tool

In this paper, a QCA designer algorithm [43] which is a well-known computational tool for sophisticated QCA circuits is used to simulate proposed circuit layouts and functionality checking and verify R-Fs in a bistable approximation [7, 34, 44, 45].

The bistable simulation engine supposes that each cell is a simple two-state system. For cell $i$ the polarization state of the cell can be shown by Eq. (3) [43].

$$P_i = \frac{\frac{E_{i,j}^k}{2\gamma}\Sigma_j P_j}{\sqrt{1+\left(\frac{E_{i,j}^k}{2\gamma}\Sigma_j P_j\right)}} \tag{3}$$

Where $P_j$ the neighboring cells polarization state and $E_{i,j}^k$ is the kink energy between cell $i$ and $j$. $\gamma$ is the tunneling energy of electrons within the cell. The summation is over all cells within an effective radius of cell $i$, and can be set prior to the simulation as shown in Fig. 6.

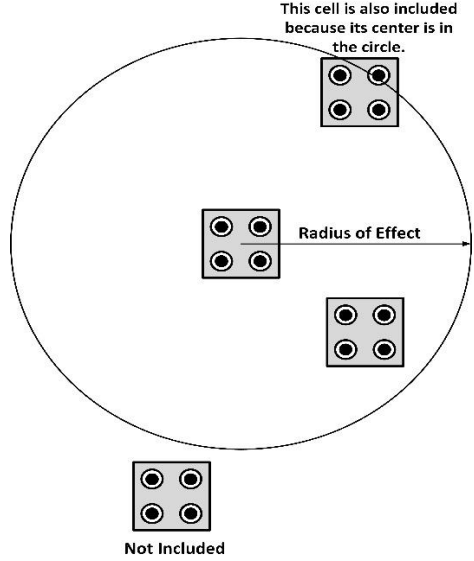

Fig. 6: Effective Radius

We define the $E_{i,j}^k$ as the Kink energy, where it represents the energy cost of cells $i$ and $j$ which own the opposite polarizations. The kink energy can be calculated, where all charges (mobile/immobile) interact between each other electrostatically. We can simply calculate the electrostatic interaction between each dot in the cell $i$ and in the cell $j$, given in Eq. (4).

$$E_{i,j}^k = \frac{1}{4\pi\epsilon_0\epsilon_r}\frac{q_i q_j}{|r_i - r_j|} \tag{4}$$

Where $\epsilon_0$ is the permittivity of free space and $\epsilon_r$ is the relative permittivity of the material system.

### 2.5. Reversible logic

In addition, static and dynamic power consumption, which are two well-known sources of power dissipation in a logic circuit, information loss is another source of energy loss in a logic circuit which was introduced by Landauer (In 1961) [3-4]. By establishing a one-to-one mapping, the reversible computation is achieved at a logical level, where the input and output vectors in the circuit are considered for the one-to-one mapping [4]. The bijective property ca be presented by (1-to-1) mapping, where in the quantum technology, in reversible systems the fan-out is not feasible. Due to lost of information in the minority input, the Majority Voting (MV) function is logically irreversible in QCA technology. The lost happens during the computations. In fact, the Bennett clocking which can be considered as another clocking arrangement in QCA can be utilized for reversible computing. Therefore, a practical realization of reversible computing can be realized by using the QCA. In the case of QCA circuits which refers to devise such as such as MV and fan-out, energy dissipation per switching event is much less than $kTln2$ and it also don't exist any physical limitation. The conclusion has been achieved by direct calculation that with Bennett clocking [2, 41]. We can conclude that the fan-out is not a necessary feature for creating RGs in QCA technology. Bennet clocking scheme has been introduced in QCA Clocking section.

#### 2.5.1. Basic reversible gates

A reversible gate (RG) realizes a reversible function. A logic function is reversible if and only if there is a one-to-one mapping between its input and output vectors [46]. Some of the essential RGs such as Fredkin gate [47], Toffoli gate [46], RUG gate [48] and Peres gate [49] are shown in Fig. 7. As shown in Fig. 7.a) Fredkin gate has a control line so that if control line A is one then, two inputs B and C are swapped in output. Fig. 7.b) show a Toffoli gate. This gate has two control lines A and B, so that if each two control lines be one then input C is inverse in output. RUG gare is shown in Fig. 7.c). This gate creates a MG in output A and in outputs B and C act as equations $AB + A'C'$ and $BC' + B'C$, respectively. Peres gate is shown in Fig. 7.d). This gate act as a Toffoli gate in output C and act as exclusive OR inputs A and B in output B. Toffoli and Fredkin gates are both universal, i.e., each logical reversible circuit can be implemented using one of these gates.

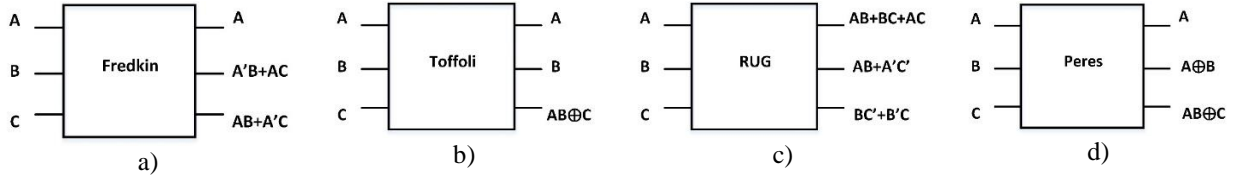

Fig. 7: RGs: a) Fredkin gate b) Toffoli gate c) RUG gate d) Peres gate.

*2.5.2. Evaluation metrics of reversible circuits*

Evaluation of the designed circuits is an important issue in the synthesis of RL circuits. to evaluate and compare different reversible designs the following metrics are used [50]:

- Number of gates can be used as a metric in circuits where gates have similar size and type. In QCA technology, MG and NG are fundamental gates. So, in this paper, number of MG and NG are used as an evaluation metric.
- Number of levels in the circuit which are required to realize the given logic functions.
- Delay that can be stated as the number of clocking zones in QCA technology.
- Number of Constant inputs and GOs. The unutilized outputs that are employed to preserve the reversibility function of the circuits are referred to GOs. The unvarying input in a reversible circuit is termed ancilla.
- Control inputs that are a new important feature that we add in this paper for evaluation metrics of reversible circuits in QCA technology. This feature is referred to as the number of AND/OR gates used in the circuit. For creating these gates using MG in QCA technology, constant inputs are necessary that this redundancy constant inputs generate extra heat.

## 3. Illustration of the present method

In this section, the proposed method is explained. First, a multi-objective synthesis method is presented for R-Fs. Then, based on this method, many well-known R-Fs are synthesized. Then for these functions, a new design of QCA layout is proposed. Finally, a new method for converting IR-FS to R-Fs is proposed.

### 3.1. Multi-objective synthesis method

In this section, by extending the synthesis flow proposed in [22], with adding an important objective to previous synthesis flow, a multi-objective synthesis method for R-Fs is presented. In this paper, in addition to objective priorities proposed in [22] which include the gate counts (the number of MGs), gate levels and the number of NGs, another objective namely Control inputs are used. This objective is considered for the first time in this paper. It's important because one of the important objectives in creating R-Fs in the QCA technology is reducing the number of AND/OR gates that need to redundancy constant inputs. In the other words, in QCA technology, AND/OR gates are generated using MG and it needs to redundancy constant inputs and because these gates are irreversible, so, according to Landauer principle [3], these redundancy constant inputs generate extra heat. So, for creating reversibility in QCA technology, the number of AND/OR gates must be reduced. Hence, in addition to objectives proposed in [22], numbers of redundancy constant inputs which are called control inputs are considered as the latest objective.

The other steps and explanations in synthesis flow are the same as proposed in [22]. Fig. 8 show this synthesis flow as rewritten for R-Fs. in this figure, MSM, denote Majority Specification Matrix proposed in [22].

```
Input: a given determination of reversible function
Output: a synthesized RG using of the MG

1.  Make the MSM given that the number of function inputs
2.  For every $f_i$ $(i = 1:m$   m: the number of function outputs)
3.  {   For every of MSMs made
4.    {  Detect the most analogous column to the feature of the main function (MF) in MSM
5.       Detect the most analogous column given that features of columns combination in MSM.
6.       Detect the most analogous feature between the MF and columns of inputs.
7.       Iterate Lines 4 & 5 for supplementary function (fi), again.
8.       Iterate Lines 4 & 5 of the algorithm for all methods explained in [22] and keep the results.

9.       Apply the method of C-K-map to the MF and keep the result.
     }
    }.
10. Utilize majority expression made in every $f_i$ for synthesis other $f_j$ $(j \neq i)$
11. Chose results between outputs to the most common expressions.
12. For decreasing the number of NGs, utilize the below specifications
    $Maj(a', b', c')' = Maj(a, b, c)$
```

Fig. 8: An algorithm of the synthesis of a reversible function

One of the most frequently used reversible logic gates (RLGs) is Fredkin gate or controlled-SWAP gate. It is $3 \times 3$ RG. For instance, in the following, the synthesis flow for the synthesis of Fredkin gate is explained. Specification function of Fredkin gate is presented in Table 1.

Table 1: Specification function Fredkin gate

| a | b | c | P | Q | R | a | b | c | P | Q | R |
|---|---|---|---|---|---|---|---|---|---|---|---|
| 0 | 0 | 0 | 0 | 0 | 0 | 1 | 0 | 0 | 1 | 0 | 0 |
| 0 | 0 | 1 | 0 | 0 | 1 | 1 | 0 | 1 | 1 | 1 | 0 |
| 0 | 1 | 0 | 0 | 1 | 0 | 1 | 1 | 0 | 1 | 0 | 1 |
| 0 | 1 | 1 | 0 | 1 | 1 | 1 | 1 | 1 | 1 | 1 | 1 |

For the synthesis of the RG by applying the synthesis flow, first, a column of outputs that is equal to corresponding input columns is separated (output column $P$ in Table 1); then other outputs according to proposed synthesis flow are synthesized ($Q$ and $R$). Table 2 and Table 3 show the synthesis of the outputs of $Q$ and $R$, respectively. First, the most similar column is chosen (c), by using a post-processing method explained in [22], specifications $F_2$ and $F_3$ are obtained as shown in Table 2 (Columns 4 and 5).

Table 2: Synthesis of output $Q$ from Fredkin gate

| a | b | c | Q | $F_1 = c$ | $F_2$ | $F_3$ |
|---|---|---|---|-----------|-------|-------|
| 0 | 0 | 0 | 0 | 0 | $X = 0$ | $X = 0$ |
| 0 | 0 | 1 | 0 | 1 | 0 | 0 |
| 0 | 1 | 0 | 1 | 0 | 1 | 1 |
| 0 | 1 | 1 | 1 | 1 | $X = 1$ | $X = 1$ |
| 1 | 0 | 0 | 0 | 0 | $X = 1$ | 0 |
| 1 | 0 | 1 | 1 | 1 | $X = 1$ | $X = 0$ |
| 1 | 1 | 0 | 0 | 0 | $X = 1$ | 0 |
| 1 | 1 | 1 | 1 | 1 | $X = 1$ | $X = 0$ |

Boolean expressions for F2 and F3 are:

$$F_2 = b + a = Maj(b, a, 1), \qquad (5)$$

$$F_3 = a'b = Maj(a', b, 0).$$

The total expression for the specification function $Q$ is:

$$Q = Maj(c, Maj(b, a, 1), Maj(a', b, 0)). \qquad (6)$$

Table 3: Synthesis of output $R$ of Fredkin gate

| a | b | c | R | $F_1 = c$ | $F_2$ | $F_3$ |
|---|---|---|---|---|---|---|
| 0 | 0 | 0 | 0 | 0 | $X = 1$ | 0 |
| 0 | 0 | 1 | 1 | 1 | $X = 1$ | $X = 0$ |
| 0 | 1 | 0 | 0 | 0 | $X = 1$ | 0 |
| 0 | 1 | 1 | 1 | 1 | $X = 1$ | $X = 1$ |
| 1 | 0 | 0 | 0 | 0 | $X = 0$ | $X = 0$ |
| 1 | 0 | 1 | 0 | 1 | 0 | 0 |
| 1 | 1 | 0 | 1 | 0 | 1 | 1 |
| 1 | 1 | 1 | 1 | 1 | $X = 1$ | $X = 1$ |

As illustrated in Table 3, firstly, the most alike column is selected (column of input $c$), then, Boolean functions of F2 and F3 are obtained as the expressions shown in (7):

$$F_2 = a' + b = Maj(a', b, 1) \qquad (7)$$

$$F_3 = ab = Maj(a, b, 0)$$

The final Boolean expression for $R$ is:

$$R = Maj(c, Maj(a', b, 1), Maj(a, b, 0)) \qquad (8)$$

The number of MG and NG are 6 and 1; thus, the number of total gates needed for the creation of Fredkin gate is 7.

Based on the above method, a new synthesis of other popular R-Fs (except Fredkin that explained in above) such as Toffoli gate, RUG, Peres and reversible full-adder (R-FA) are presented in Table 6. In this table, underlines show fanouts. Also, for synthesizing R-FA, specification functions introduced in [21] are used.

### 3.2. Converting irreversible to reversible functions

In this sub-section, a new method for converting irreversible IR-Fs to R-Fs is proposed in the QCA technology. For this purpose, first, common states in outputs are signed, then, input and output columns are compared with each other which can remove the most common states in outputs selected and as a new column in the output added. This work will be repeated till common states in outputs fully deleted and the one to one mapping between the outputs and inputs is created. Utilizing this method, each kind of R-Fs with the minimum intricacy, GOs, control inputs and delay, area can directly be created while the other methods such as [51-53] have used popular RBs like Toffoli and Fredkin gates or building of RBs (intermediate RBs) to design other R-Fs in which these methods are not optimal. Finally, the obtained reversible function is synthesized using the proposed method in Section 3.1.

Proposition: by using the above method a reversible function with the lowest number of MG and GOs is created.

Proof:

Suppose the number of function inputs is $n$, thus there are $2^n$ states in the truth table where in each of columns, half ($\frac{2^n}{2}$) of states are ones and another half ($\frac{2^n}{2}$) are zeros. Then, the following two cases may be occurred for the function outputs.

1. If the number of 1s in output is not equal to $\frac{2^n}{2}$, so at least $n$ inputs should be added to them to create reversibility in the output. As a reversible function needs $2^n$ distinct states in the output. On the other hand, the number of GO and input data are identical. As a result, no additional MG needs to be created.

2. If the number of 1s in the output is equal to $\frac{2^n}{2}$, then reversibility can be obtained by $n - 1$ garbages such that these extra outputs are not equal to input orders. Thus, these outputs lead to extra MGs. In addition, extra MGs can be deleted if GOs are equal to function outputs; but this leads to common states created in the output without reversibility. But reversibility can be obtained when $n$ inputs are added to the output. Then, a reversible function is generated with the lowest number of MGs and GOs.

### 3.3. The 13 standard combinational functions

For verification, the proposed method is applied to the 13 standard combinational functions introduced in [23] which are initially utilized for comparison purposes. These functions represent all 256 three-variable Boolean functions. First, the 13 standard combinational functions must be converted to R-Fs. Finally, the obtained reversible function is synthesized using the proposed method in Section 3.1. To comparative study, the number of gates and clocking zones of the 13 standard functions which are implemented to use each of the four RGs by [54] are reported in Table 4. It points to the fact that the realization of a logic circuit using the proposed method can result in better cost-effective designs than those of with other conventional RGs in terms of the numbers of majority, NGs and clocking zone. The obtained circuits of 13 standard functions displayed in Table 4 are illustrated in the appendix.

Table 4: Reversible implementation of 13 standard functions

| Functions | Fredkin+Inverter | | | Toffoli+Inverter | | | QCA 1 | | | QCA 2 | | | Proposed Method | | |
|---|---|---|---|---|---|---|---|---|---|---|---|---|---|---|---|
| | Majority | NOT | Clk zone | Majority | NOT | Clk zone | Majority | NOT | Clk zone | Majority | NOT | Clk zone | Majority | NOT | Clk zone |
| F1=AB'C | 12 | 2 | 9 | 8 | 4 | 9 | 6 | 4 | 5 | 6 | 4 | 5 | 2 | 1 | 3 |
| F2=AB | 6 | 1 | 4 | 4 | 2 | 4 | 3 | 2 | 2 | 3 | 2 | 2 | 1 | 0 | 1 |
| F3=A'BC+A'B'C' | 12 | 3 | 9 | 8 | 5 | 9 | 6 | 4 | 5 | 6 | 4 | 5 | 3 | 3 | 4 |
| F4=A'BC+AB'C' | 12 | 2 | 9 | 12 | 7 | 9 | 9 | 6 | 8 | 9 | 6 | 8 | 5 | 3 | 4 |
| F5=A'B+BC' | 12 | 2 | 9 | 8 | 4 | 9 | 6 | 4 | 5 | 6 | 4 | 5 | 2 | 1 | 3 |
| F6=AB'+A'BC | 12 | 2 | 9 | 12 | 6 | 9 | 9 | 6 | 5 | 9 | 6 | 5 | 4 | 2 | 4 |
| F7=A'BC+ABC'+A'B'C' | 18 | 4 | 9 | 12 | 8 | 9 | 9 | 6 | 5 | 9 | 6 | 5 | 4 | 3 | 4 |
| F8=A | 6 | 1 | 4 | 4 | 2 | 4 | 3 | 2 | 2 | 3 | 2 | 2 | 0 | 0 | 1 |
| F9=AB+AC+BC | 18 | 4 | 9 | 16 | 8 | 14 | 3 | 2 | 2 | 3 | 2 | 2 | 1 | 0 | 2 |
| F10=A'B+B'C | 6 | 1 | 4 | 12 | 6 | 9 | 9 | 6 | 5 | 9 | 6 | 5 | 3 | 2 | 3 |
| F11=A'B+BC+AB'C' | 18 | 4 | 9 | 4 | 2 | 4 | 12 | 8 | 5 | 12 | 8 | 5 | 4 | 3 | 4 |
| F12=AB+A'B' | 6 | 2 | 4 | 4 | 2 | 4 | 6 | 4 | 5 | 6 | 4 | 5 | 3 | 2 | 3 |
| F13=ABC'+A'B'C'+AB'C+A'BC | 12 | 4 | 9 | 8 | 5 | 9 | 6 | 4 | 5 | 6 | 4 | 5 | 3 | 3 | 3 |

### 3.4. New design of QCA layout of reversible functions

In this sub-section, new designs of QCA cell layouts for popular RLGs are shown in Fig. 9. These layouts are related to RGs synthesized in the above sections including Toffoli, Fredkin, R-FA, RUG, and Peres, respectively.

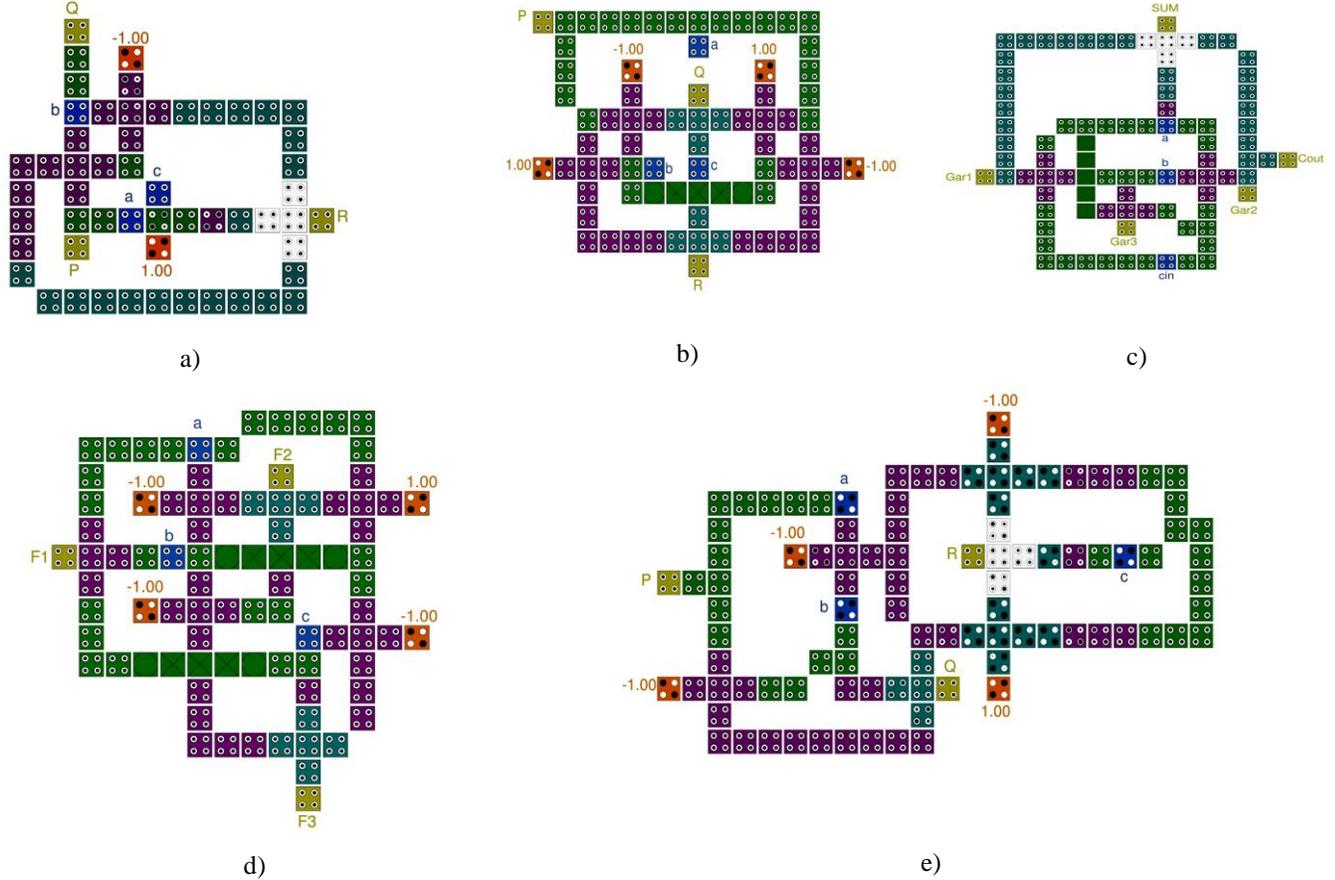

Fig. 9: QCA cell layouts of RLGs. a) Toffoli gate b) Fredkin gate c) reversible full adder d) RUG gate e) Peres gate.

## 4. Results

### 4.1. Comparison results

As illustrated in Table 6, the proposed method is compared with the best existing obtained results for popular RGs in logical level. In this table, in addition to the number of MG and NG and levels, the number of Constant inputs is considered as an objective. Synthesized circuits are shown in the final column of this table. As shown in this table, the proposed approach results in fewer MGs, NGs, Constant inputs, and fewer logic levels are compared to available methods [18-22]. Also, reduction of the number of Control inputs as an objective in algorithm has led to reduction of the numbers of AND/OR gates and it results in the reduction of power consumption. In this table, for synthesizing R-FA, specification function introduced in [21] is used. In addition, a detailed comparison between the proposed implementation and the best presented implementations is performed in Table 7 from different aspects which include area, cell amount and the number of clocking zones. Clearly, our designs outperform the proposed designs in [18–22] with respect to the area, cell amount and the number of clocking zones in the design of QCA layouts.

### 4.2. Simulation results

As stated in Section 2.4, in this paper, QCA designer is used for simulation of the proposed designs. Simulator version used in this study is 2.0.3. Numeric values of simulation parameters are summarized in Table 5. Also, simulation results related to the new design of QCA layouts of R-Fs include Toffoli, Fredkin, RUG, Peres and R-FA

(shown in Fig. 9) are shown in Fig. 10.

Table 5: A QCA designer settings

| Parameter | Value | Parameter | Value |
|---|---|---|---|
| Size of cell | 18 nm | clock high | 9.8e−22 J |
| Samples number | 12800 | clock low | 3.8e−23 J |
| Tolerance of convergence | 0.001000 | clock amplitude factor | 2.000 |
| Effect radius | 65 nm | layer separation | 11.5000 nm |
| Relative permittivity | 12.9 | Number of maximum iterations per sample | 100 |

Table 6: Comparisons with popular R-Fs

| RGs | Method | Constant inputs | Number of NGs | Number of MGs | Level | Synthesized circuit |
|---|---|---|---|---|---|---|
| Toffoli | [18] | 4 | 2 | 4 | 3 | P=a,<br>Q=b,<br>R=Maj(Maj(a,b,c')',Maj(a,c,1),Maj(b,c',0)) |
| | Proposed method | 2 | 2 | 4 | 2 | |
| Fredkin | [20] | 6 | 3 | 6 | 2 | P=a,<br>Q=Maj(c,Maj(b,a,1),Maj(a',b,0)),<br>R=Maj(c,Maj(a',b,1),Maj(a,b,0)) |
| | Proposed method | 4 | 1 | 6 | 2 | |
| RUG | [17] | 6 | 3 | 7 | 2 | f1=M(a,b,c),<br>f2=M(c',M(a',b,1),M(a,b,0)),<br>f3=M(c,M(b,c',0),M(b,c,0)') |
| | Proposed method | 4 | 3 | 7 | 2 | |
| Peres | [19] | 7 | 4 | 7 | 3 | P=a,<br>Q=M(b,M(a,b,0)',M(a,b',0)),<br>R=M(M(M(a,b,0),c',0),M(M(a,b,0)',c',1),c) |
| | Proposed method | 4 | 3 | 6 | 3 | |
| R-FA | [21] | 0 | 4 | 6 | 2 | Sum=M(M(a,b,cin)',a,M(a',b,cin)),<br>Cout=M(a,b,cin),<br>Gar1=M(a',b,cin),<br>Gar2=Cout,<br>Gar3=M(a,b,cin') |
| | Proposed method | 0 | 3 | 4 | 2 | |

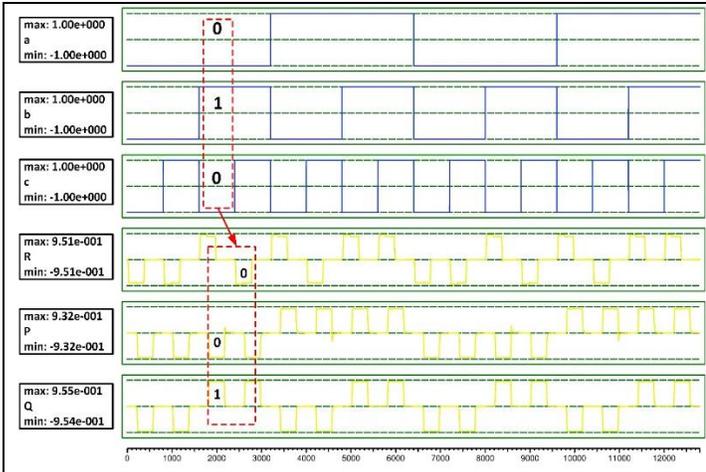

a) Toffoli gate

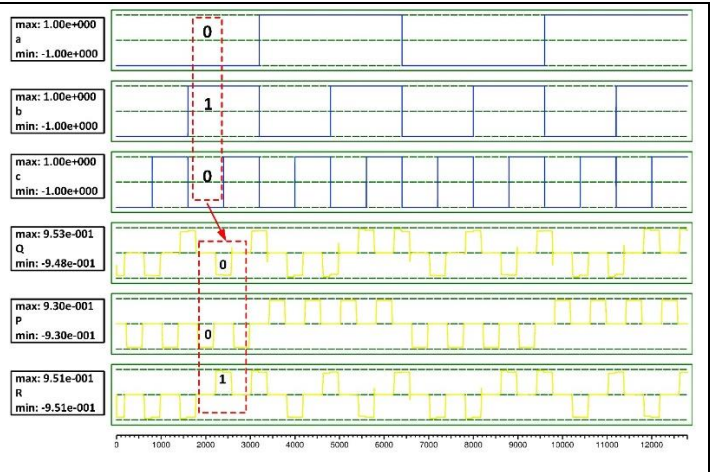

b) Fredkin gate

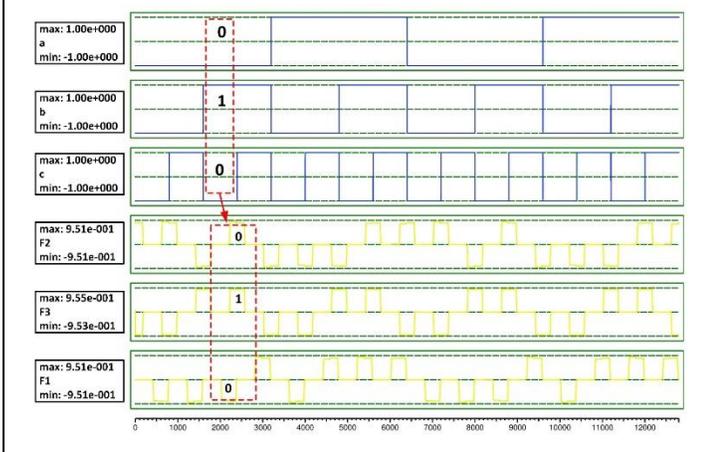

c) RUG gate

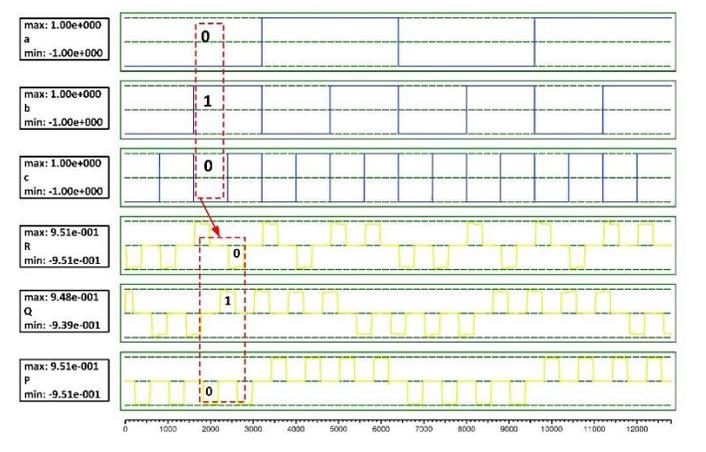

d) Peres gate

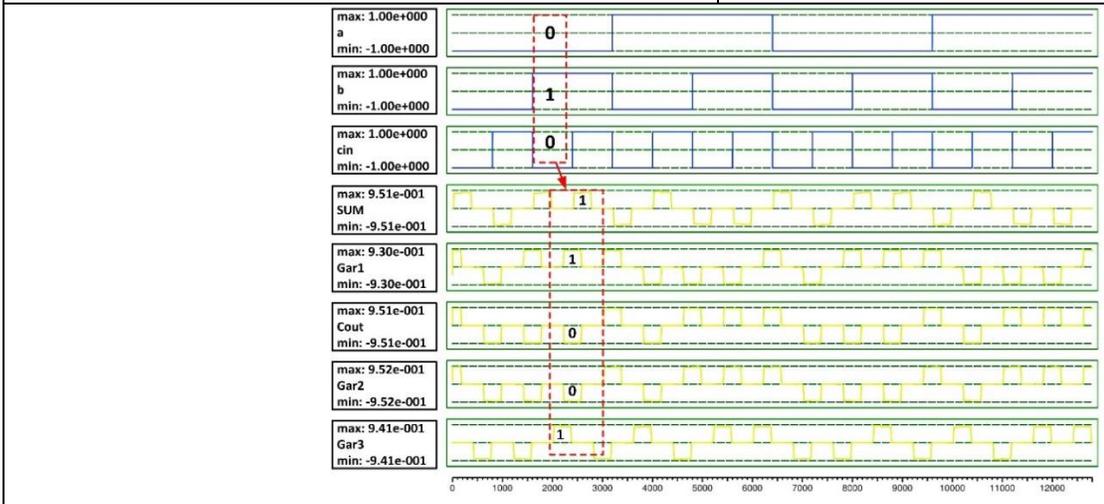

e) R-FA

Fig. 10: Simulation results of RGs in Fig. 6

Table 7: Performance comparison of QCA implemented RGs

| RGs | Method | Cell amount | Area ($\mu m^2$) | Clocking zones |
|---|---|---|---|---|
| Toffoli | [18] | 48 | 0.067 | 4 |
| | Proposed Method | 54 | 0.060 | 4 |
| Fredkin | [20] | 93 | 0.087 | 3 |
| | Proposed Method | 82 | 0.09 | 3 |
| RUG | [17] | ~298 | 0.37 | 7 |
| | Proposed Method | 97 | 0.10 | 3 |
| Peres | [19] | 99 | 0.10 | 4 |
| | Proposed Method | 98 | 0.13 | 4 |
| R-FA | [21] | 371 | 0.33 | 6 |
| | Proposed Method | 93 | 0.13 | 4 |

## 5. Conclusion

In this paper, the main contribution is classified in three parts. In the first part, by extending the synthesis method proposed in [22], a multi-objective synthesis method was proposed for R-Fs, gate levels, the number of NGs and control inputs which were the input cells with fixed polarization used for programming 2-input OR and AND gates). The reduction of control inputs was a new objective that was used in this paper for improving the reduction of power consumption in R-Fs. Based on the proposed method, a new synthesis of many of familiar RGs such as Toffoli, Fredkin and so on was proposed. In the second part, a new method for converting IR-FS to R-Fs in QCA technology was presented. This innovative approach has some advantages over the available methods in the literature including direct and optimal conversion of an irreversible function to its counterpart reversible function. The method introduced in this study does not need to call for any optimization method as a subroutine (e.g. sub-optimal method of using conventional RBs such as Toffoli and Fredkin are eliminated). Because of the minimum number of GOs, the GOs do not generate any external gate (e.g. MG and NG). For showing efficiency of the proposed method, it is applied to the 13 standard combinational functions proposed in [23], and a reversible function was created with the lowest number of MG and GOs. Finally, in the Third part, new designs of QCA layouts were presented for gates synthesized in the previous section. Results showed that our proposed method outperforms the most efficient methods with respect to area, complexity (cell amount), delay (clocking zones), and in logic level with respect to levels, control inputs, number of MG and NG.

**Appendix**

In this section, the obtained circuits of 13 standard functions shown in Table 4 are illustrated in Table A1.

Table A1: the obtained circuits of 13 standard functions illustrated in Table 4

| Functions | Synthesized circuit |
|---|---|
| F1=AB'C | F1=Maj(Maj(A,B',0),C,0), Y2=B, Y3=C, Y4=A, |
| F2=AB | F2=Maj(A,B,0), Y2=B, Y3=C, Y4=A |
| F3=A'BC+A'B'C' | F3=Maj(Maj(A',B,C'),Maj(A',B',C),0), Y2=B, Y3=C, Y4=A |
| F4=A'BC+AB'C' | F4=Maj(Maj(Maj(A',B,0),C,0),Maj(Maj(A,B',0),C',0),1), Y2=B, Y3=C, Y4=A |
| F5=A'B+BC' | F5=Maj(Maj(A,C,0)',B,0), Y2=B, Y3=C, Y4=A |
| F6=AB'+A'BC | F6=Maj(Maj(A,B,C),Maj(A,B',0),Maj(A',B',1)), Y2=B, Y3=C, Y4=A |
| F7=A'BC+ABC'+A'B'C' | F7=Maj(Maj(A',B,C'),Maj(A',B',C),Maj(A,C',0)),Y2=B, Y3=C, Y4=A |
| F8=A | F8=A |
| F9=AB+AC+BC | F9=Maj(A,B,C),Y2=B, Y3=C, Y4=A |
| F10=A'B+B'C | F10=Maj(C,Maj(A',B',1),Maj(B,C',0)),Y2=B, Y3=C, Y4=A |
| F11=A'B+BC+AB'C' | F11=Maj(Maj(A,B,C'),Maj(B',C',0),Maj(A',B,C)),Y2=B, Y3=C, Y4=A |
| F12=AB+A'B' | F12=Maj(B,Maj(B',A,1),Maj(A',B',0)),Y2=B, Y3=C, Y4=A |
| F13=ABC'+A'B'C'+AB'C+A'BC | F13=Maj(Maj(A,B',C'),B,Maj(A',B',C)),Y2=B, Y3=C, Y4=A |